\documentclass[12pt]{iopart}
\usepackage{indentfirst}
\usepackage{harvard}
\usepackage{graphicx}
\expandafter\let\csname equation*\endcsname\relax
\expandafter\let\csname endequation*\endcsname\relax
\usepackage{amsmath}

\begin{document}
\bibliographystyle{jphysicsB}

\title[Visualization of FC layer weights in DL CT reconstruction]{Visualization of fully connected layer weights in deep learning CT reconstruction}

\author{Qiyang Zhang$^{1,2}$, Dong Liang$^{1,2}$}

\address{$^1$ Paul C Lauterbur Research Center for Biomedical Imaging, Research Center for Medical Artificial Intelligence, Shenzhen Institutes of Advanced Technology, Chinese Academy of Sciences, Shenzhen, Guangdong 518055, China}
\address{$^2$ Paul C Lauterbur Research Center for Biomedical Imaging, Shenzhen Institutes of Advanced Technology, Chinese Academy of Sciences, Shenzhen, Guangdong 518055, China}
\ead{qy.zhang@siat.ac.cn,dong.liang@siat.ac.cn}

\date{\today}

\begin{abstract}
\noindent
Recently, the use of deep learning techniques to reconstruct computed tomography (CT) images has become a hot research topic, including sinogram domain methods, image domain methods and sinogram domain to image domain methods. All these methods have achieved favorable results. In this article, we have studied the important functions of fully connected layers used in the sinogram domain to image domain approach. First, we present a simple domain mapping neural networks. Then, we analyze the role of the fully connected layers of these networks and visually analyze the weights of the fully connected layers. Finally, by visualizing the weights of the fully connected layer, we found that the main role of the fully connected layer is to implement the back projection function in CT reconstruction. This finding has important implications for the use of deep learning techniques to reconstruct computed tomography (CT) images. For example, since fully connected layer weights need to consume huge memory resources, the back-projection function can be implemented by using analytical algorithms to avoid resource occupation, which can be embedded in the entire network.
\end{abstract}
\maketitle

\section{Introduction}
Currently, neural networks and deep learning have completely changed the traditional way of signal processing, image processing, image recognition, etc. Many of these successful cases have been applied to the medical field \cite{krishnan2016genome,gulshan2016development,jin2017deep,chen2017low,wang2018image,kang2018deep,yang2018low,zhang2018sparse}. 
Also, the authors of a recent article \cite{wang2016perspective} described the vision of using machine learning to create new CT image reconstruction algorithms to improve conventional analysis and iterative methods. Now, more and more academic research hotspots for CT reconstruction are focused on DL (deep learning) methods, including sinogram domain methods \cite{lee2018deep,pelt2013fast}, image domain methods \cite{han2018framing,shan20183} and sinogram domain to image domain methods \cite{wurfl2018deep,zhu2018image}. In particular, the domain transformation method has great potential to remove noise and artifacts simultaneously during graphics reconstruction \cite{zhu2018image}. 
However, there is currently no clear mathematical explanation for each part of the domain mapping network. In this article, we use visualization techniques to study the function of the fully connected layer used in the domain mapping network.

\section{Methodology}
\subsection{Prepare and train the simple network}
\begin{figure}[h]
\centering
\includegraphics[width=0.75\textwidth]	{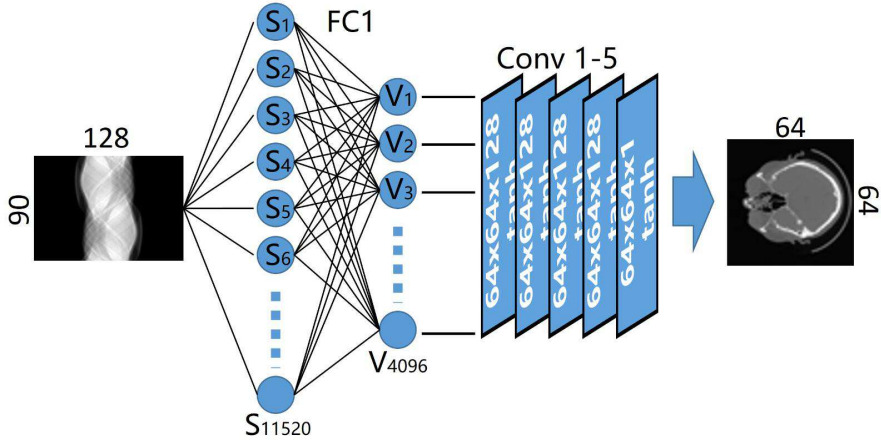}
\caption{The architecture of the simple network. The network with only one fully connected layer followed by five convolutional layers, each convolutional layer with 128 filters, except for the final layer that has only 1 filter. The weight kernels for all convolutional layers are $3\times3$ with a filter stride 1. The three numbers in each box denotes the image column number, image row number and the channel number, respectively. The activation function for all layers is tanh.}
\label{fig:simple_network}
\end{figure}

In order to explain what the fully connected layer of the network, like AUTOMAP \cite{zhu2018image} for CT reconstruction, learned, we built a simple network, as illustrated in Fig.~\ref{fig:simple_network}, with only one fully connected layer followed by five convolutional layers, each convolutional layer with 128 filters, except for the final layer that has only 1 filter. a $3\times3$ filter with a filter stride 1 is used for all convolutional layers. The activation function for all layers is tanh. The loss function was a simple squared loss between the network output and target image intensity values. 
Here we only want to explain the role of the fully connected layer, so all used CT images size were downsampled to $64\times64$. A fan-beam CT imaging geometry was simulated for this simple network. The source to detector distance was 1500.00 mm, and the source to rotation center was 1000.00 mm. There were 128 detector elements, each had a dimension of 1.6 mm. To make the CT images fit in this simulated imaging geometry, we further assumed that all CT images have the same pixel dimension of $1.0 mm\times1.0 mm$. Forward projections, i.e., Radon projections, were collected from 90 views with 4.00 degree angular interval. Notice that this is only a simulated fan-beam CT imaging geometry to to explain the role of the fully connected layer. Also be aware that no noise is added in these simulations.

As illustrated in Fig.~\ref{fig:simple_network}, the shape of the input image, which be fed to fully connected layer, needs to be reshaped from two dimension ($90\times128$) to one dimension($11520$). Also, the output shape of the fully connected layer(FC1), which be fed to CNNs, needs reverse change from one dimension(($4096$)) to two dimension($64\times64$) . 

The network was trained by Adam algorithm \cite{kingma2014adam} with starting learning rate of  $10^{-5}$. The learning rate was exponentially decayed by a factor of 0.96 after every 1000 steps. The mini-batch had a size of 60, and batch-shuffling was turned on to increase the randomness of the training data. The network was trained for 200 epochs on the Tensorflow deep learning framework using a single graphics processing unit (GPU, NVIDIA GeForce GTX 1080Ti) with 11 GB memory capacity.

\begin{figure}[h]
\centering
\includegraphics[width=0.55\textwidth]{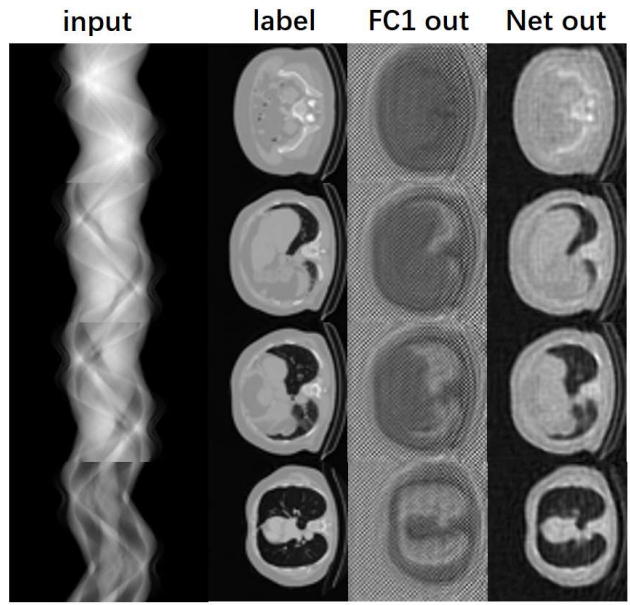}
\caption{Numerical experimental results of simple neural network.}
\label{fig:simp_net_rst}
\end{figure}

The numerical experimental results of this simple neural network are shown in Fig.~\ref{fig:simp_net_rst}. Sinogram image is fed to this network and CT reconstruction image be generated from it. Because we only using this simple neural network to analyze what the fully connected layer has learned, so wo not do quantitative analysis. But the mid result of this network, which was reshaped for human-readable, was shown(FC1 out). As  shown in Fig.~\ref{fig:simp_net_rst} FC1 out is somehow close to label. 

\subsection{Change the shape of the fully connetced layer weights}
In order to observe the physical meaning of the weights learned by the network, we need to recombine the shape of the full connection layer weights. As illustrated in Fig.~\ref{fig:simple_network}, where $S_i$ denote the reshaped projection data. $V_j$ denote the output of fully connected layer. $W_{i,j}$ is refers to one weight of FC1 (fully connected layer), which denote from input end $i$ to output end $j$. There are $11520\times4096$ weights.

The reshape rules from two dimension ($90\times128$) to one dimension ($11520$) of input sinogram image $I$ can be defined by
\begin{equation}
S_i = I_{p,q}, \quad where\ i = (p-1) \times Q+ q , \quad 1\leq p \leq P;1\leq q \leq Q
\end{equation}
where $I_{p,q}$ represent the pixel value of  the p-th row and q-th column of the input sinogram image, $P$ and $Q$ represent the total pixels along the row and column directions. And, physically, $P$ and $Q$ represent the the total acquisition angles and detector cells. Here $P=90$ and $Q=128$.

Because of  the convolutional layers, which followed FC1, are end-to-end mapping network, So, the output data shape of FC1 layer is the same as the label image. Then the reshape rules of output of FC1 layer from one dimension ($4096$) to two dimension ($64\times64$) can be defined by
\begin{equation}
M_{c,t} = V_j, \quad where\ j=(c-1)\times T+ t, \quad 1\leq c \leq C;1\leq t \leq T
\end{equation}
where $M_{c,t}$ represent the value of  the p-th row and q-th column of the input two dimension feature data for convolutional layers, $C$ and $T$ represent the total numbers along the row and column directions of the feature data. Here $C=64$ and $T=64$.

\begin{figure}[h]
\centering
\includegraphics[width=0.75\textwidth]	{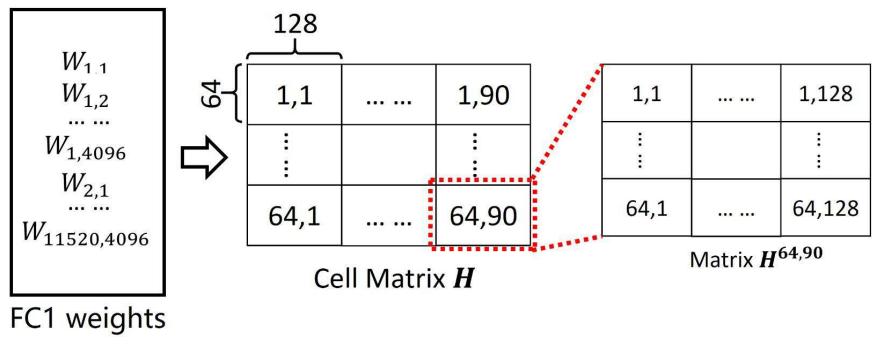}
\caption{The reshape rules of fully connected layer weights}
\label{fig:fully_connected_layer}
\end{figure}

Following the rules for changing shapes described above, we can transform the shape of the fully connected layer weights to match the rules of back projection in CT reconstruction. Because in this simulated fan-beam CT imaging geometry the sinogram datas were collected from 90 views with 4.00 degree angular interval, the detector has 128 cells and the CT image be projected with two dimension of $64\times 64$. So we can define a cell matix $H$ with dimension of $K\times L$, as shown in Fig.~\ref{fig:fully_connected_layer} where $K$ denotes total columns of the CT image and $L$ denotes total  collected views. Here $K=64$ and $L=90$. Each cell has the dimension of $A\times B$, where $A$ denotes total rows of the CT image and $B$ denotes total numbers of detector cells. Here $A=64$ and $B=128$. 

We note that $H^{k,l}$ to represent each element of the cell matix $H$ and $H^{k,l}_{a,b}$ to represent each element of cell $H^{k,l}$. So, $H^{k,l}_{a,b}$ can represent each element of the fully connected layer weights:
\begin{equation}
\begin{aligned}
H^{k,l}_{a,b} = W_{i,j}, \quad where\ i &= (k-1)\times A+(a-1) \\
j &= (l-1)\times B+(b-1),\\
1\leq k \leq K; & 1\leq l \leq L;  1\leq a \leq A;1\leq b \leq B
\end{aligned}
\end{equation}

\section{Visualization of weights}
In this section, we only see what weights map (feature map) look like, not do quantitative analysis. So, the display windows of the images shown in this section are not the same but be adjusted comfortable to see feature shape. 

\begin{figure}[h]
\centering
\includegraphics[width=0.95\textwidth]{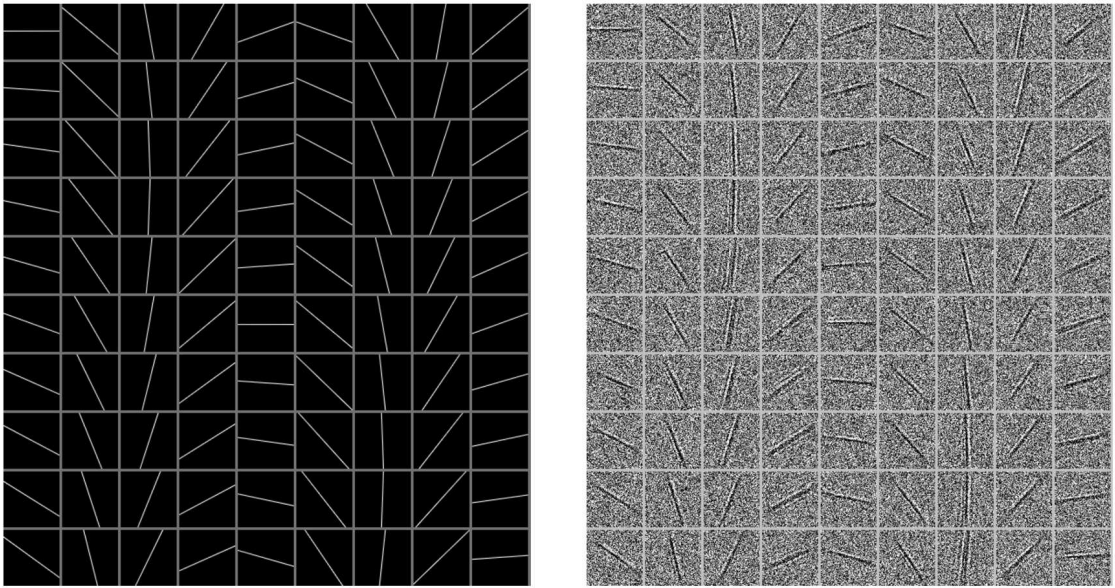}
\caption{Weights map of the 64th detector unit for different pixel values of the final reconstructed CT image at different acquisition angles. Left weights map is calculated by analytic algorithm under the same rules as the right fully connected layer weights map. Both map images from top to bottom and from left to right are the 0th back projection view weights, the 2th, ..., the 90th back projection view weights.}
\label{fig:visl_fix_unit_diff_deg}
\end{figure}

First, let's visualize the weights of one given fixed detector unit for different pixel values of the final reconstructed CT image at different acquisition angles. In order to make it more obvious, we have selected the 64th detector unit ($H^{k,l}_{a,b},b=64$). Let ($H^{k,l}_{a,b},b=64;l=1;k=1;2,...,K;a=1,2,...,A$) be the first ($l=1$) back projection view of CT reconstruction weights map, ($H^{k,l}_{a,b},b=64;l=2;k=1;2,...,K;a=1,2,...,A$) be the second ($l=2$) ,...,($H^{k,l}_{a,b},b=64;l=90;k=1;2,...,K;a=1,2,...,A$) be the last ($l=90$) back projection view of CT reconstruction weights map. The displayed of those maps are show on the right image of Fig.~\ref{fig:visl_fix_unit_diff_deg}. The image on the left of Fig.~\ref{fig:visl_fix_unit_diff_deg} is weights map directly calculated by the back projection analytic algorithm at the same rule of above.

\begin{figure}[h]
\centering
\includegraphics[width=0.95\textwidth]{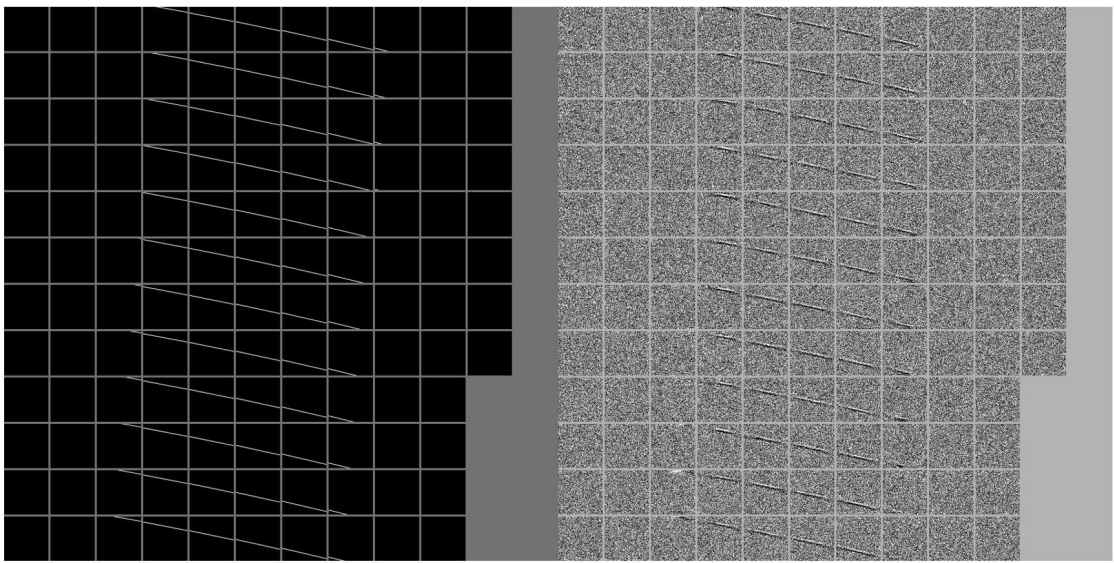}
\caption{Weights map of different detector unit for different pixel values of the final reconstructed CT image at a fixed acquisition angle,12th degree. Left weights map is calculated by analytic algorithm under the same rules as the right weights map. Both map images from top to bottom and from left to right are the 0th detector unit weights at the 12th back projection view of CT reconstruction, the 2th, ..., the 128th detector unit weights at the 12th back projection view of CT reconstruction.}
\label{fig:visl_fix_deg_diff_unit}
\end{figure}

Second, let's visualize the weights of different detector unit for different pixel values of the final reconstructed CT image at a fixed acquisition angle. Here we select the 12th view degree for visualization ($H^{k,l}_{a,b},l=12$), of course, we can also choose other degrees.
Let ($H^{k,l}_{a,b},b=1;l=12;k=1;2,...,K;a=1,2,...,A$) be the first ($b=1$) first detector unit weight map at the 12th back projection view of CT reconstruction, ($H^{k,l}_{a,b},b=2;l=12;k=1;2,...,K;a=1,2,...,A$) be the second ($b=2$) ,...,($H^{k,l}_{a,b},b=128;l=12;k=1;2,...,K;a=1,2,...,A$) be the last ($b=128$) detector unit weight map at the 12th back projection view of CT reconstruction. The displayed of those images are show on the right image of Fig.~\ref{fig:visl_fix_deg_diff_unit}. The image on the left of Fig.~\ref{fig:visl_fix_deg_diff_unit} is weights map directly calculated by the back projection analytic algorithm at the same rule of above.

From Fig.~\ref{fig:visl_fix_unit_diff_deg} and Fig.~\ref{fig:visl_fix_deg_diff_unit}, we can realize that, the feature map of fully connected layer of this trained network is the same as that of analytic algorithm, they are very sparse data. The fully connected layer implements mapping from sinogram domain to image domain and do some filtering. In Fig.~\ref{fig:simp_net_rst}, FC1 out image appears to be inversely numerical in comparison with the label image or the net output image. This phenomenon can also be seen from Fig.~\ref{fig:visl_fix_unit_diff_deg} and Fig.~\ref{fig:visl_fix_deg_diff_unit}. Because of the back propagation process of deep learning affects all weight parameters, include fully connected layer weights and convolutional layer weights, so, this phenomenon (inversely numerical and some filted) affected weights may be learned out at the fully connected layer . 
After the above analysis, we can draw a conclusion that the main role of the fully connected layer is to implement the back projection function in sinogram domain to image domain mapping CT reconstruction networks. 

\section{Discussion}
The main problem of using neural networks to realize back-projection is that it consumes huge memory resources. Just like this simple network, it can only be used to reconstruct CT images of $64\times64$ size, the number of weights in fully connected layer was 45M, if we use 4Bytes for one float type data, the consumption of memory resources is 180M Bytes. If we use the clinical CT image size of $512\times512$ at 360 collection views, 768 detector units, the number of weights in one fully connected layer will reach more than 69G. With today's computer graphics storage technology, it is impossible to achieve such a large amount of data storage in high-speed computing. Therefore, it is currently impossible to obtain practical applications.

By visualizing the weights of fully connected layers, it shows that back-projection can be learned through neural networks. This means that we can directly use analytic algorithms to implement back projection, so that the network can learn more complex problems, such as noise reduction and artifact suppression.

\section*{References}
\bibliography{myscript4FullyCF}

@Article{jin2017deep,
  author    = {Jin, Kyong Hwan and McCann, Michael T and Froustey, Emmanuel and Unser, Michael},
  title     = {Deep convolutional neural network for inverse problems in imaging},
  journal   = {IEEE Transactions on Image Processing},
  year      = {2017},
  volume    = {26},
  number    = {9},
  pages     = {4509--4522},
  publisher = {IEEE},
}

@Article{chen2017low,
  author    = {Chen, Hu and Zhang, Yi and Kalra, Mannudeep K and Lin, Feng and Chen, Yang and Liao, Peixi and Zhou, Jiliu and Wang, Ge},
  title     = {Low-dose {CT} with a residual encoder-decoder convolutional neural network},
  journal   = {IEEE transactions on medical imaging},
  year      = {2017},
  volume    = {36},
  number    = {12},
  pages     = {2524--2535},
  publisher = {IEEE},
}

@Article{han2018framing,
  author    = {Han, Yoseob and Ye, Jong Chul},
  title     = {Framing {U-Net} via deep convolutional framelets: Application to sparse-view {CT}},
  journal   = {IEEE transactions on medical imaging},
  year      = {2018},
  volume    = {37},
  number    = {6},
  pages     = {1418--1429},
  publisher = {IEEE},
}

@Article{wang2018image,
  author  = {Wang, Ge and Ye, Jong Chu and Mueller, Klaus and Fessler, Jeffrey A},
  title   = {Image Reconstruction is a New Frontier of Machine Learning.},
  journal = {IEEE transactions on medical imaging},
  year    = {2018},
  volume  = {37},
  number  = {6},
  pages   = {1289--1296},
}

@Article{kang2018deep,
  author    = {Kang, Eunhee and Chang, Won and Yoo, Jaejun and Ye, Jong Chul},
  title     = {Deep convolutional framelet denosing for low-dose {CT} via wavelet residual network},
  journal   = {IEEE transactions on medical imaging},
  year      = {2018},
  volume    = {37},
  number    = {6},
  pages     = {1358--1369},
  publisher = {IEEE},
}

@Article{yang2018low,
  author    = {Yang, Qingsong and Yan, Pingkun and Zhang, Yanbo and Yu, Hengyong and Shi, Yongyi and Mou, Xuanqin and Kalra, Mannudeep K and Zhang, Yi and Sun, Ling and Wang, Ge},
  title     = {Low dose {CT} image denoising using a generative adversarial network with Wasserstein distance and perceptual loss},
  journal   = {IEEE transactions on medical imaging},
  year      = {2018},
  publisher = {IEEE},
}

@Article{zhang2018sparse,
  author  = {Zhang, Zhicheng and Liang, Xiaokun and Dong, Xu and Xie, Yaoqin and Cao, Guohua},
  title   = {A Sparse-View CT Reconstruction Method Based on Combination of DenseNet and Deconvolution.},
  journal = {IEEE transactions on medical imaging},
  year    = {2018},
  volume  = {37},
  number  = {6},
  pages   = {1407--1417},
}

@article{lee2018deep,
  title={Deep-neural-network based sinogram synthesis for sparse-view CT image reconstruction},
  author={Lee, Hoyeon and Lee, Jongha and Kim, Hyeongseok and Cho, Byungchul and Cho, Seungryong},
  journal={arXiv preprint arXiv:1803.00694},
  year={2018}
}

@Article{zhu2018image,
  author    = {Zhu, Bo and Liu, Jeremiah Z and Cauley, Stephen F and Rosen, Bruce R and Rosen, Matthew S},
  title     = {Image reconstruction by domain-transform manifold learning},
  journal   = {Nature},
  year      = {2018},
  volume    = {555},
  number    = {7697},
  pages     = {487},
  publisher = {Nature Publishing Group},
}

@article{pelt2013fast,
  title={Fast tomographic reconstruction from limited data using artificial neural networks},
  author={Pelt, Daniel Maria and Batenburg, Kees Joost},
  journal={IEEE Transactions on Image Processing},
  volume={22},
  number={12},
  pages={5238--5251},
  year={2013},
  publisher={IEEE}
}

@article{shan20183,
  title={3-D Convolutional Encoder-Decoder Network for Low-Dose CT via Transfer Learning From a 2-D Trained Network},
  author={Shan, Hongming and Zhang, Yi and Yang, Qingsong and Kruger, Uwe and Kalra, Mannudeep K and Sun, Ling and Cong, Wenxiang and Wang, Ge},
  journal={IEEE transactions on medical imaging},
  volume={37},
  number={6},
  pages={1522--1534},
  year={2018},
  publisher={IEEE}
}

@Article{kingma2014adam,
  author  = {Kingma, Diederik P and Ba, Jimmy},
  title   = {Adam: A method for stochastic optimization},
  journal = {arXiv preprint arXiv:1412.6980},
  year    = {2014},
}

@article{krishnan2016genome,
  title={Genome-wide prediction and functional characterization of the genetic basis of autism spectrum disorder},
  author={Krishnan, Arjun and Zhang, Ran and Yao, Victoria and Theesfeld, Chandra L and Wong, Aaron K and Tadych, Alicja and Volfovsky, Natalia and Packer, Alan and Lash, Alex and Troyanskaya, Olga G},
  journal={Nature neuroscience},
  volume={19},
  number={11},
  pages={1454},
  year={2016},
  publisher={Nature Publishing Group}
}

@article{gulshan2016development,
  title={Development and validation of a deep learning algorithm for detection of diabetic retinopathy in retinal fundus photographs},
  author={Gulshan, Varun and Peng, Lily and Coram, Marc and Stumpe, Martin C and Wu, Derek and Narayanaswamy, Arunachalam and Venugopalan, Subhashini and Widner, Kasumi and Madams, Tom and Cuadros, Jorge and others},
  journal={Jama},
  volume={316},
  number={22},
  pages={2402--2410},
  year={2016},
  publisher={American Medical Association}
}

@article{wurfl2018deep,
  title={Deep learning computed tomography: Learning projection-domain weights from image domain in limited angle problems},
  author={W{\"u}rfl, Tobias and Hoffmann, Mathis and Christlein, Vincent and Breininger, Katharina and Huang, Yixin and Unberath, Mathias and Maier, Andreas K},
  journal={IEEE transactions on medical imaging},
  volume={37},
  number={6},
  pages={1454--1463},
  year={2018},
  publisher={IEEE}
}

@article{wang2016perspective,
  title={A perspective on deep imaging},
  author={Wang, Ge},
  journal={arXiv preprint arXiv:1609.04375},
  year={2016}
}
\end{document}